\documentclass[sigconf,screen]{acmart}

\setcopyright{none}
\renewcommand\footnotetextcopyrightpermission[1]{}

\usepackage{amsmath}
\usepackage{graphicx}
\usepackage{textcomp}
\usepackage{xcolor}
\usepackage{booktabs}
\usepackage{array}
\usepackage{placeins}
\AtBeginDocument{\hypersetup{urlcolor=black}}
\graphicspath{{./}{../figures/}{../charts/}{figures/}{charts/}}
\AtEndDocument{\nobalance}

\begin{document}

\title{Making Fragmented Reports Legible: Finding Patterns and Perceptions of Sexual Violence in Bangladesh}
\date{September 21, 2026}

\author{Sheherjan Haq}
\orcid{0009-0009-5024-8403}
\affiliation{%
  \institution{University of Dhaka}
  \city{Dhaka}
  \country{Bangladesh}
}
\email{sheherjanhaq@gmail.com}

\author{Sharifa Sultana}
\orcid{0000-0003-2906-7391}
\affiliation{%
  \institution{University of Illinois Urbana-Champaign}
  \city{Urbana--Champaign}
  \country{USA}
}
\email{sharifas@illinois.edu}

\begin{abstract}
Evidence about sexual violence in Bangladesh is fragmented across individual reports, while official and civil-society statistics rarely provide reusable case-level detail. We examine how structured analysis can make one part of this fragmented record legible without treating it as prevalence data. Our corpus contains 2,811 articles timestamped 2013--2023 from the Prothom Alo publishing ecosystem; 2,794 include parseable metadata about reported victims, alleged perpetrators, incidents, legal responses, and locations. We combine descriptive and spatial analysis of these records with thematic analysis of 115 convenience-sample survey responses collected in late 2020. The corpus documents many young, female, and student victims, frequent acquaintance and neighbor relationships, uneven geographic documentation, and substantial missingness in legal outcomes. Respondents most often describe weak enforcement, insecurity, patriarchal socialization, education gaps, and community inaction as conditions enabling persistence. These findings characterize news documentation and public perceptions; they do not estimate incidence, geographic risk, or causality. We contribute an uncertainty-aware framing for HCI research using sensitive, low-resource news data and identify design requirements for provenance, privacy, validation, and responsible communication.
\end{abstract}



\maketitle

\section{Introduction}
Sexual violence remains one of the most urgent and unresolved social crises in Bangladesh. Although public outrage, media attention, and legal action often follow individual rape incidents, similar cases continue to recur across the country. This repetition shows a critical limitation of case-by-case response: it reacts to visible incidents but does not fully explain the patterns and conditions that allow such violence to persist. A broader understanding requires examining where rape cases are reported more frequently, how they appear across victims, perpetrators, time, location, and legal action, and what social or institutional conditions may accompany their recurrence. Yet existing evidence is scattered across news reports, legal summaries, survivor accounts, surveys, and localized studies, making it difficult to connect individual cases into a larger picture. Computational work has shown that digital narratives can support structured characterization of harassment and sexual-violence reports, while also exposing reporting silence and bias~\cite{karlekar2018safecity,garrett2019silence,klemmer2021delays}. HCI research further cautions that systems concerning gendered violence must foreground survivor privacy, safety, and stakeholder context~\cite{matthews2017survivors,freed2017ipv}. Community perceptions are also important because they reveal how people understand the causes of rape, the failures of prevention, and the forms of legal, educational, and social action they consider necessary.

Existing studies on sexual violence in Bangladesh remain constrained by the scarcity of large, structured, and accessible data. Many incidents are not available in digital form, and collecting, cleaning, locating, and organizing case details requires substantial effort. Bangladesh Mahila Parishad maintains an important civil-society monitoring practice by compiling annual counts from national newspapers; for example, its 2021 report drew on 13 dailies~\cite{bmp2021report}. This work documents visibility and scale but is generally disseminated as organizational reporting rather than a reusable, case-level scientific dataset. The closest computational Bangladesh study analyzed 10,191 rape-news headlines posted by ten media outlets on Facebook from 2013 to 2021, identifying victim and judicial-process topics but not constructing the incident-, legal-, and geographic-level representation pursued here~\cite{alzaman2023headlines}. Survivor-centered studies reveal trauma, stigma, reporting barriers, and the burden placed on victims, but they mostly focus on consequences after violence has occurred. Another body of work studies specific spaces where women face risk. Workplace-focused studies show how gendered power relations, job insecurity, and weak institutional safeguards expose women to abuse at work. Mobility-focused studies on transport safety highlight harassment and insecurity in buses, streets, and commuting spaces. Legal-system studies examine reporting, investigation, and prosecution barriers, showing why justice is often delayed or weakened, but not why similar cases continue to recur across places and communities.

This gap motivates three questions that guide our study:

\begin{itemize}
    \item \textbf{Where:} How is geographic documentation distributed across available division, district, and subdistrict tags?
    \item \textbf{How:} How do these cases appear across victims, attackers, timing, incident context, and legal action?
    \item \textbf{Perceived why:} What social, cultural, and institutional conditions do surveyed community members associate with the persistence of sexual violence?
\end{itemize}

To answer these questions, we analyze 2,811 articles from the Prothom Alo publishing ecosystem. The records include article narratives, publication timestamps, URLs, location tags, and a structured metadata field; 2,794 metadata objects are parseable. We examine victim profiles, alleged-perpetrator characteristics, incident context, location, and reported legal response. We further analyze 115 convenience-sample community survey responses to describe perceived enabling conditions and prevention priorities. This combination connects two kinds of evidence while keeping their epistemic roles separate: news records describe what one publisher documented, whereas survey responses describe what this respondent group believed.

Using district and subdistrict tags, we map the distribution of documented reports. These raw counts reveal where this corpus contains more tagged articles, not where violence is more prevalent or risk is higher. Beyond location, the metadata describe predominantly young female victims; among records with known occupation or status, students form the largest group. Alleged perpetrators are often described as acquaintances or neighbors as well as strangers. Community responses associate persistence with weak law enforcement, gendered socialization, insecurity, and limited economic empowerment. We use these patterns to motivate follow-up questions and design implications rather than causal conclusions.

The main contributions of this paper are as follows:

\begin{itemize}
    \item We characterize a structured, case-oriented corpus derived from fragmented reporting in one Bangladeshi news ecosystem, including its severe temporal, geographic, and field-level missingness.
    \item We show how spatial views can expose documentation concentration while explicitly communicating why raw media counts must not be presented as geographic risk.
    \item We connect reported-case attributes with community perceptions without treating either source as prevalence or causal evidence.
    \item We derive HCI requirements for provenance, extraction validation, privacy, uncertainty display, and responsible use of sensitive low-resource-language data.

\end{itemize}

\section{Related Work}
Understanding sexual violence in Bangladesh requires an interdisciplinary view that connects lived experience with cultural, institutional, and socioeconomic conditions. Studies of convicted perpetrators, public-space harassment, and structural stigma suggest overlapping influences rather than isolated individual deviance~\cite{jahan2025perpetrators,berik2024men,scheer2020stigma}. Research therefore draws on multiple incomplete data streams to compensate for the scarcity of comprehensive official statistics. This motivates a geographically sensitive synthesis, but it does not make causal inference possible from observational news and perception data.

Studies outside Bangladesh show several complementary ways to analyze reported sexual violence. In India, SafeCity narratives were annotated and computationally classified to recover forms of harassment and details useful for incident reporting~\cite{karlekar2018safecity}; critical discourse analysis of Indian newspaper coverage shows how language and media structure shape the public meaning of rape~\cite{nagar2016reporting}. In Sweden, a study of newspaper reporting from 1990--2015 identified recurring genres that could reproduce stereotypes while also shaping debate and local planning~\cite{nilsson2019rape}. In the United States, spatial models of police data found substantial geographic variation in rape-reporting delays and explicitly treated reporting as a process distinct from occurrence~\cite{klemmer2021delays}. A Brazilian study similarly used population-normalized notification rates and spatial statistics rather than raw counts to study intimate-partner rape~\cite{silva2024brazil}. These studies make the methodological contrast clear: media text can reveal documentation and framing, whereas geographic risk claims require defensible denominators, reporting-process assumptions, and validated event data.

The empirical corpus divides broadly into three complementary strands. First, qualitative work, including reflexive thematic analyses, in-depth interviews, focused ethnographies, and discourse studies, has been indispensable for surfacing lived experience, institutional failures, and media framings that normalize aggression. Studies of rape perpetrators document distorted understandings of consent, denial of responsibility, orthodox gender norms, adverse childhood experiences, poor education, and financial hardship among convicted offenders~\cite{jahan2025perpetrators}. Related studies of intra-familial rape, children experiencing sexual violence, and survivor burden of proof further show how family relations, consent confusion, rape myths, and disbelief constrain recognition and reporting~\cite{rahman2024children,wiecek2024intrafamilial,guajardo2024burden}. Second, quantitative and mixed-methods research provides statistical grounding for root-cause hypotheses: surveys, case-control designs, purposive sampling of vulnerable occupational groups, and spatially informed questionnaires have linked autonomy, poverty, social isolation, informal employment, and population density to elevated risk and under-reporting~\cite{stoff2021ipv,ibrahim2025autonomy,huq2023sexworkers,emon2025transport}. Third, archival, media-content, and digital-platform analyses extend the temporal and discursive reach of the field by tracing policy implementation gaps, historical patterns of institutional neglect, online harassment, and the rhetorical construction of women in public discourse~\cite{mowri2023framing,nova2019online,saha2024computing}.

Workplace and public-space studies are especially relevant for Bangladesh because women's mobility and employment often expose them to overlapping economic and gendered vulnerabilities. Research on the ready-made garments (RMG) sector shows that gender-based violence and harassment are embedded in weak labor regulation, managerial silence, precarious employment, low bargaining power, and patriarchal workplace cultures~\cite{akter2024rmg,basirulla2024garmentroot,siddiquee2025threads}. Studies of public transport in Dhaka similarly connect harassment to overcrowding, insufficient transport supply, weak law enforcement, limited family support, and the everyday negotiation of risk by women commuters~\cite{mowri2023framing,mowri2024thesis,emon2025transport}. Broader urban-safety research further suggests that women's route choices are shaped by lighting, crowd movement, visibility, media-generated fear, and perceived availability of help~\cite{lino2024walk}. These findings indicate that sexual violence cannot be separated from spatial design, labor markets, institutional accountability, and the politics of respectability.

Health and psychosocial studies add another layer by showing how violence both reflects and produces long-term vulnerability. Research on rural Bangladesh finds that social connection, especially instrumental support, may reduce intimate partner violence severity~\cite{stoff2021ipv}, while national evidence links women's autonomy with lower approval of intimate partner violence~\cite{ibrahim2025autonomy}. Studies also connect sexual violence with suicide risk in Bangladesh and with long-term depression and self-reported mental and physical health consequences after childhood sexual assault~\cite{arafat2021suicide,akinyemi2025longterm}. Campus-focused research on substance use, victim-status awareness, and reporting barriers adds a comparative lens, showing how normalization of assault and uncertainty about resources can suppress reporting even outside the Bangladeshi context~\cite{schwarz2017campus}.

Despite methodological variety, the literature exhibits persistent gaps relevant to a survey of perceived enabling conditions. Longitudinal evidence is scarce, limiting causal inference about temporal trends and policy effects. Geographic coverage is uneven and heavily Dhaka-centric, with comparatively little district-level or multi-city analysis for rural--urban comparison. Health impacts are also rarely synthesized with sociological, occupational, digital, and spatial evidence. Mechanistic links between media objectification, anonymous online harassment, migration status, transport exclusion, and perpetration remain under-explored~\cite{mowri2023framing,nova2019online,scheer2020stigma}. Finally, many mixed-methods studies lack explicit integration protocols.

Policy and intervention literature emphasizes integrated service delivery, workplace protections, public awareness, and transport safety, but evaluations frequently omit how corruption, overpopulation, and local governance deficits impede implementation. One-Stop Crisis Centres and multidisciplinary case management models show promise for improving reporting, respectful care, and survivor satisfaction, yet evidence on long-term outcomes and scalability in high-density informal settlements is limited~\cite{moonajilin2026occ}. Prevention-oriented proposals in the literature tend to focus on education, sectoral enforcement, and awareness campaigns; comparatively little attention has been paid to digital emergency tools designed for low-connectivity, low-literacy contexts or to upstream measures that directly target structural drivers such as economic precarity and governance failures.

\section{Methodology}
This study follows a convergent descriptive design that places structured analysis of media-reported metadata beside quantitative and qualitative summaries of survey responses. The methodological goal is not to estimate prevalence, geographic risk, temporal change, or causality. Instead, we characterize what is documented in the available corpus and compare those patterns with community perceptions.

\subsection{Data Sources}
The first dataset, \textit{updated\_rape\_case\_meta\_data\_version\_2}, contains 2,811 article rows. Of the URLs, 2,791 are on the main Prothom Alo domain and the remaining 20 are on four affiliated Prothom Alo domains. Each row includes a headline, publication timestamp, article content, URL, optional division/district/subdistrict tags, and a nested metadata field. A total of 2,794 metadata objects are parseable. The metadata describe victim age, gender and status; alleged-perpetrator count and relationship; incident characteristics; police and complaint fields; legal outcome; and geographic tags. We use the article row as the unit of analysis because the repository does not contain a documented incident-deduplication key; this may count multiple reports about one event and is treated as a validity threat.

Coverage is highly uneven over time: only eight rows precede 2018, 161 are from 2018, and 2,642 (94.0\%) are from 2019--2023. We therefore show the full distribution as a source-coverage diagnostic, not a longitudinal trend. Geographic tags are also incomplete: 35.7\% of rows lack division, 26.7\% lack district, and 52.5\% lack subdistrict. Spatial figures use only records with the relevant tag and always display raw documentation counts.

The second dataset is the \textit{Bangladesh Rape Case Survey}, which contains 115 responses collected between November 22 and December 9, 2020. The survey includes respondent gender, closed-ended questions on belief in a rape-free society and nearby exposure to abuse, and open-ended questions on perceived causes, prevention, and community action. The repository does not document the recruitment channel, sampling frame, eligibility rules, consent procedure, or respondent geography. We therefore treat it as a small convenience sample and do not generalize percentages to Bangladesh.

\subsection{Data Preparation}
The case metadata were parsed from nested text fields into tabular variables. Bengali division, district, and subdistrict labels were normalized for plotting and spatial joining. Numeric fields such as victim age and number of alleged attackers were converted into analyzable formats. Missing and unknown values were retained rather than interpreted as negative evidence. The current repository does not preserve the system, prompt/schema creation process, or human validation used to produce the nested metadata; accordingly, all extracted-field results are provisional until a stratified manual validation is completed.

The survey data were cleaned by normalizing clear response variants in closed-ended questions. Open-ended responses were grouped provisionally around repeated concepts such as legal reform, education, social awareness, patriarchal norms, social security, family responsibility, and bystander action. Keyword-mention counts are presented only as descriptive navigation aids: they are not independent thematic-analysis results and do not measure the prevalence or importance of a belief. A submission-ready analysis requires a documented codebook, at least two coders, disagreement resolution, and reflexive reporting.

\begin{figure}[tbp]
\centering
\includegraphics[width=\linewidth]{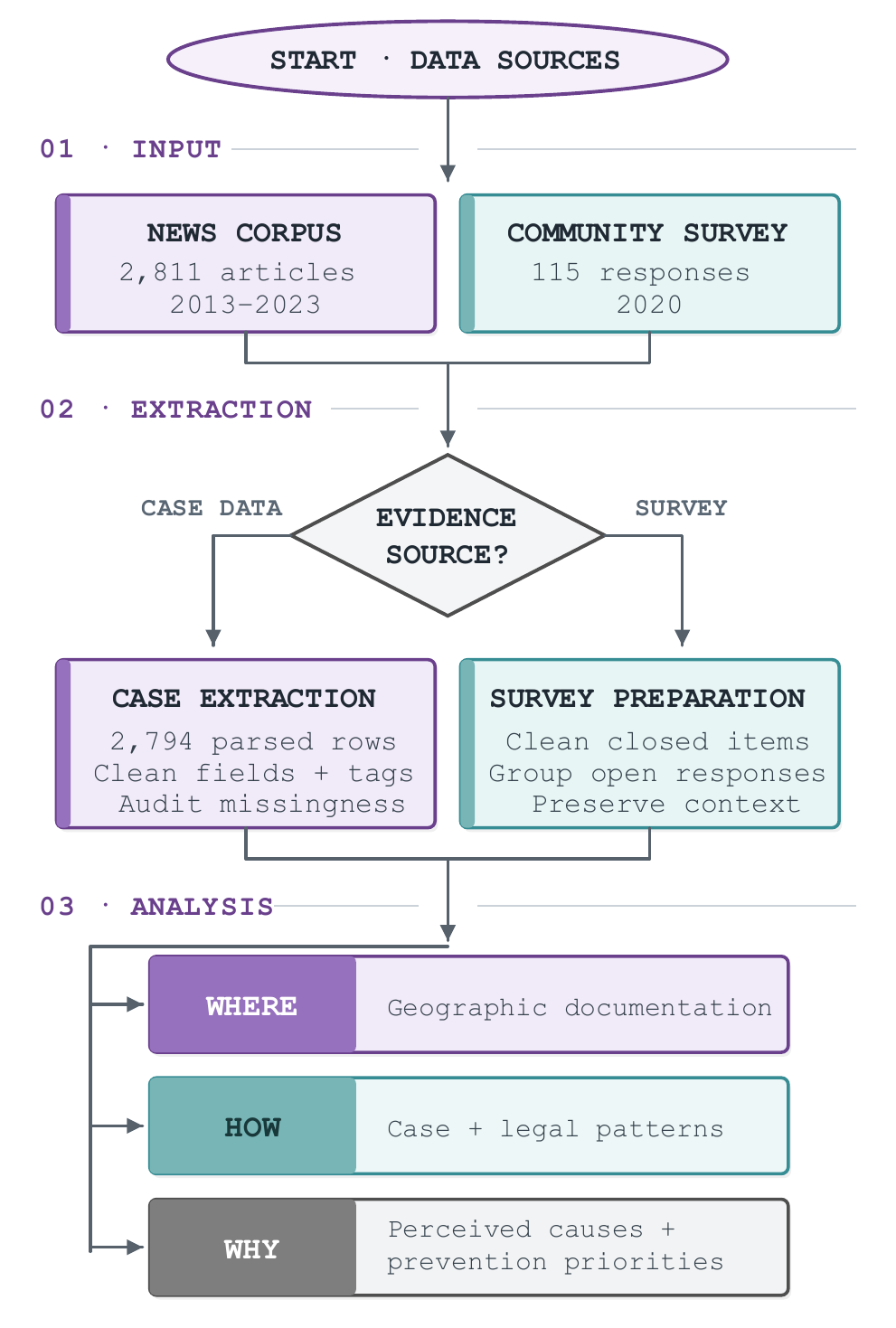}
\caption{Study workflow from two evidence sources through source-specific preparation and three parallel analyses. The reported-case corpus supports the \textit{where} and \textit{how} analyses, while the survey supports the \textit{why} analysis of perceived causes and prevention priorities.}
\Description{A vertical flowchart begins with two inputs: 2,811 Prothom Alo--ecosystem article records and 115 community survey responses. A diamond routes the sources into separate extraction and preparation processes. The prepared data feed three parallel analysis rectangles labeled Where, How, and Why.}
\label{fig:methodology-flowchart}
\end{figure}

Fig.~\ref{fig:methodology-flowchart} summarizes the workflow from the two data sources through source-specific preparation and the where, how, and why analyses.

\subsection{Analytical Procedure}
The analysis proceeded in four stages. First, we profiled corpus provenance, temporal coverage, location-tag completeness, and field missingness. Second, we summarized available victim, alleged-perpetrator, incident, and legal-response fields with explicit denominators. Third, we mapped raw documentation counts at division, district, and subdistrict levels; no population adjustment or risk estimation was performed. Fourth, we summarized closed survey items and provisionally grouped open responses. We compared the two data streams in a joint interpretive step, looking for convergence and tension while avoiding causal triangulation.

\subsection{Ethical Considerations}
The study uses secondary media-derived records and anonymized survey data. The analysis avoids publishing names, exact residences, URLs tied to findings, or case-level narratives. This matters because the source metadata retain names and residences and can create re-identification and contextual-privacy risks even when the underlying articles are public. Survey quotations are shortened and labeled only by broad respondent attributes; distinctive quotations should undergo an additional searchability review before publication. We treat the corpus as evidence of what was reported and recorded, not national incidence, and survey findings as perceptions rather than verified causal claims. Formal ethics-review/consent documentation is not present in the repository and must be supplied before submission.

\section{Findings}
We present the findings in two connected layers. First, the article metadata show what this publishing ecosystem documented about temporal coverage, victims, alleged perpetrators, incident context, geography, and legal response. Second, the survey shows how a small group of respondents understood persistence, responsibility, and prevention. These layers are interpreted together but not collapsed: article metadata are not prevalence data, and respondent explanations are not verified causes.

The main text therefore retains the two complete victim and alleged-perpetrator profile tables alongside a small set of figures. Appendix~\ref{app:detailed-case-tables} now provides the supporting coverage, incident, and legal-response audits without duplicating those central profile tables.

\subsection{Temporal Coverage}
\begin{figure}[htbp]
\centerline{\includegraphics[width=\linewidth]{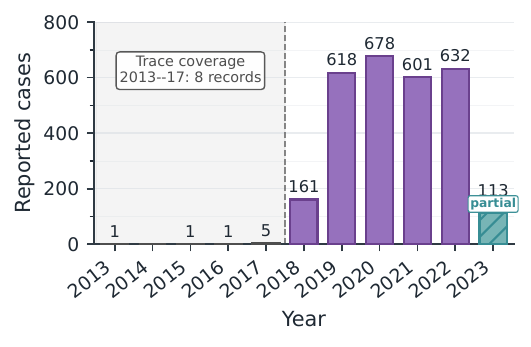}}
\caption{Publication-year coverage of all 2,811 acquired Prothom Alo--ecosystem article records. The 2013--2017 interval contains only eight records; 2018 spans January 2--December 31, whereas 2023 spans only January 2--April 10. The bars describe corpus acquisition, not incidence or change in sexual violence.}
\Description{A bar chart covering 2013 through 2023. The gray, lightly shaded 2013--2017 region is labeled trace coverage and contains eight records in total. Purple bars show 2018 through 2022. A hatched teal 2023 bar contains 113 records and carries a small partial label; the caption gives the exact coverage dates.}
\label{fig:cases-year}
\end{figure}

The corpus contains 2,811 article rows, of which 2,794 have parseable structured metadata. Publication timestamps span 2013--2023, but the coverage is not longitudinally balanced: 2013--2017 contain only eight rows in total, 2018 contains 161, and 2019--2023 contain 2,642. Fig.~\ref{fig:cases-year} is therefore a corpus-coverage diagnostic. It cannot support claims that sexual violence increased or decreased over time, and the low 2023 count may reflect incomplete acquisition.

The imbalance is substantial: 2,529 records (90.0\% of all acquired rows) were published during 2019--2022, whereas the five years before 2018 contribute only eight. The 2018 records cover the full calendar year, but 2023 stops on April 10. Consequently, aggregate patterns in the remainder of this section are driven mainly by the 2019--2022 documentation window. Comparisons of yearly totals, seasonality, or pre/post change would confound article acquisition with whatever change might have occurred in reporting or violence.

\newcommand{\fieldcoveragetable}{%
\begin{table}[tbp]
\centering
\caption{Selected field-coverage audit. Percentages use the 2,794 parsed records except the first row, which uses all 2,811 acquired rows.}
\label{tab:field-coverage}
\footnotesize
\renewcommand{\arraystretch}{1.10}
\begin{tabular}{@{}p{0.43\linewidth}rr@{}}
\toprule
\textbf{Field} & \textbf{Available} & \textbf{Missing} \\
\midrule
Parseable metadata & 2,794 (99.4\%) & 17 \\
Victim age & 1,626 (58.2\%) & 1,168 \\
Victim condition/status & 1,582 (56.6\%) & 1,212 \\
District tag & 2,043 (73.1\%) & 751 \\
Subdistrict tag & 1,317 (47.1\%) & 1,477 \\
Legal outcome & 1,216 (43.5\%) & 1,578 \\
\bottomrule
\end{tabular}
\end{table}
}

\subsection{Victim Profiles}
Victim metadata identify the victim as female in 2,596 of 2,794 parsed records (92.9\%). Age is available in 1,626 records, with a median of 15 years. Three extracted ages fall outside the 1--80 plotting range; among the remaining 1,623 records, the 13--17 and 6--12 groups contain 574 and 434 records, respectively (Fig.~\ref{fig:case-patterns}a). These are patterns of visibility in the acquired articles, not population estimates. Among specified condition/status labels, student is by far the largest category (906 records; Fig.~\ref{fig:case-patterns}b).

Victim status is known in 2,425 records: 2,087 are marked alive and 338 deceased. The latter value describes the status documented in the acquired article, not a mortality rate, because articles may have been collected at different stages and follow-up reporting is not linked. Marital status is absent in 1,811 records. Among the 983 specified values, married (462) and single (450) are similar in count; divorced (46) and widowed (25) are much less frequent. Education is missing in 1,977 records; among the 817 specified values, secondary (428) and primary education (241) dominate.

Together, these fields make a strong documentation pattern visible: many records concern girls, adolescents, and students. That pattern cannot be generalized to all victim-survivors in Bangladesh because news selection may favor cases involving children or students, and metadata extraction can recover only details present in recognizable language. Table~\ref{tab:victim-demo} presents the full victim profile, including field availability and omitted values, directly in the findings.

\newcommand{\victimdemotable}{%
\begin{table}[tbp]
\centering
\caption{Victim demographic patterns in the parsed case metadata.}
\label{tab:victim-demo}
\footnotesize
\renewcommand{\arraystretch}{1.18}
\begin{tabular}{@{}>{\centering\arraybackslash}m{0.34\linewidth}|>{\raggedright\arraybackslash}m{0.58\linewidth}@{}}
\toprule
\textbf{Field} & \textbf{Result} \\
\midrule
\textbf{Parsed records} &
2,794 usable metadata records from 2,811 raw rows. \\
\midrule
\textbf{Victim gender} &
Female: 2,596 (92.9\%).\newline
Male: 71 (2.5\%).\newline
Unknown: 127 (4.5\%). \\
\midrule
\textbf{Victim age} &
Available: 1,626 records.\newline
Median: 15 years.\newline
Mean: 17.1 years.\newline
Age 13--17: 574 records.\newline
Age 6--12: 434 records. \\
\midrule
\textbf{Victim status} &
Alive: 2,087 (86.1\% of known).\newline
Deceased: 338 (13.9\% of known).\newline
Unknown: 369 records. \\
\midrule
\textbf{Victim condition} &
Student: 906.\newline
Housewife: 194.\newline
Garment worker: 86.\newline
Speech-impaired: 62.\newline
Other specified: 334.\newline
Unknown: 1,212. \\
\midrule
\textbf{Marital status} &
Married: 462 (47.0\% of known).\newline
Single: 450 (45.8\% of known).\newline
Divorced: 46 (4.7\% of known).\newline
Widowed: 25 (2.5\% of known).\newline
Unknown: 1,811. \\
\midrule
\textbf{Education level} &
Secondary: 428 (52.4\% of known).\newline
Primary: 241 (29.5\% of known).\newline
Higher secondary: 80 (9.8\% of known).\newline
Graduate: 62 (7.6\% of known).\newline
Other specified: 6.\newline
Unknown: 1,977. \\
\bottomrule
\end{tabular}
\end{table}
}

\victimdemotable

Fields such as family economic status, physical injuries, and psychological support were sparsely reported in the parsed metadata and are therefore treated as data-quality limitations rather than primary demographic indicators.

\subsection{Geographic Patterns}
\begin{figure}[htbp]
\centering
\includegraphics[width=\linewidth]{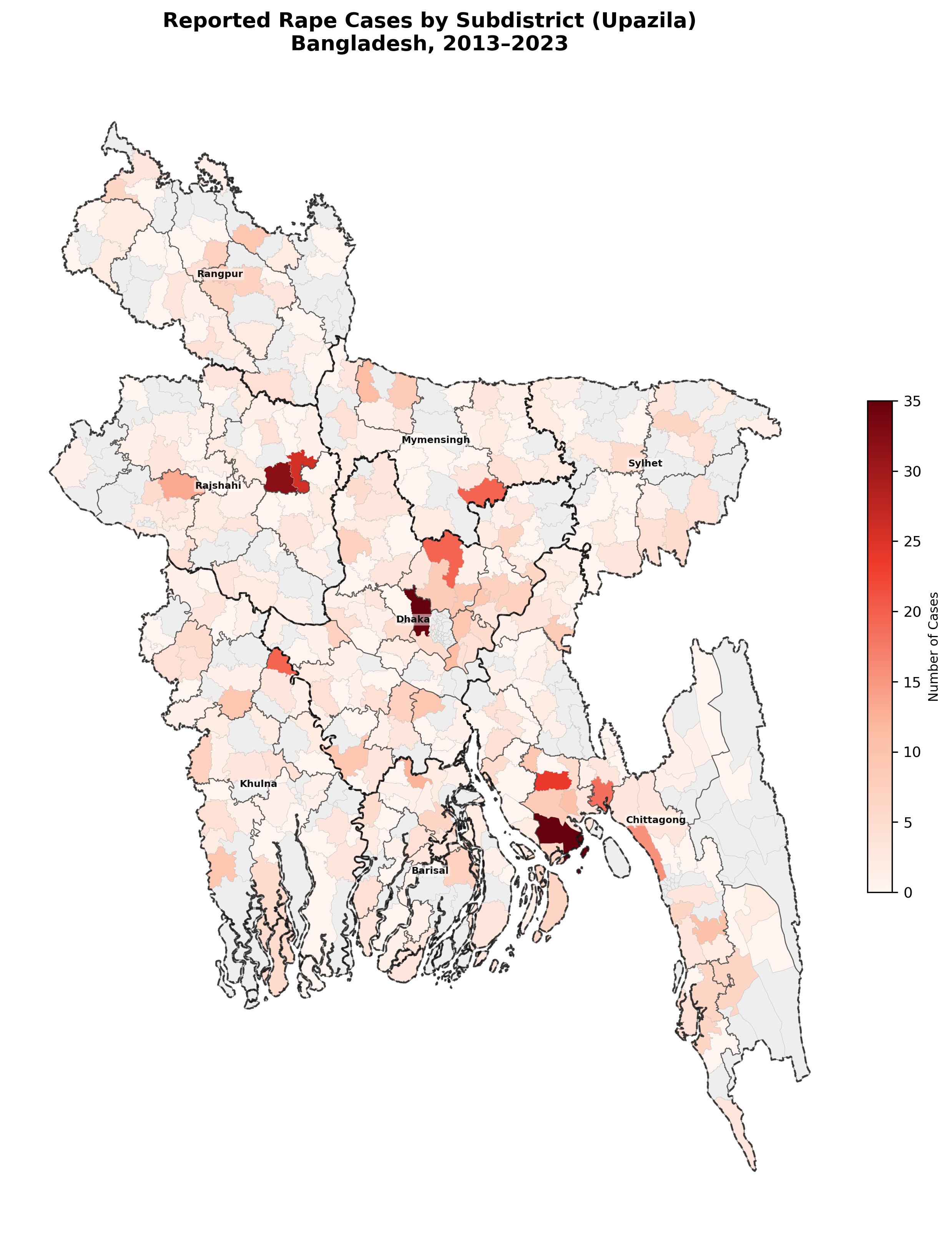}
\caption{Distribution of subdistrict-tagged article records. Darker shading indicates documentation density among the 1,317 records with a usable subdistrict tag, not population-adjusted incidence or risk. The broader district map appears in Appendix~\ref{app:district-map}.}
\Description{A choropleth map of Bangladesh at subdistrict level. Darker shading denotes more articles with a matching subdistrict tag; it does not denote population-adjusted incidence or risk.}
\label{fig:spatial-maps}
\end{figure}

The parsed metadata contain labels from all eight divisions, 64 districts, and 352 subdistricts, but label breadth is not representative coverage. District tags are absent from 751 rows (26.7\%) and subdistrict tags from 1,477 (52.5\%). Among tagged articles, Chattogram, Dhaka, Noakhali, Sylhet, and Narayanganj are frequent district labels. Fig.~\ref{fig:spatial-maps} visualizes this documentation concentration. Differences may reflect population, violence, reporting behavior, newsroom attention, search/acquisition choices, or tagging accuracy; these factors cannot be separated with the present data.

The contrast between district and subdistrict completeness is important. Nearly three quarters of parsed records have a district tag, but fewer than half have a subdistrict tag. We retain the finer subdistrict view in the main text because it makes within-district documentation concentrations visible, while the appendix district map provides the broader and more complete acquisition audit. The presence of all 64 district labels shows geographic breadth, but it does not establish uniform search recall, newsroom reach, or metadata quality across those districts.

Accordingly, darker areas should be read as places that are more visible in this corpus. They may reflect larger populations and reporting infrastructures as much as incident frequency. Low-count or untagged areas may reflect limited reporting, inconsistent place names, or acquisition gaps. The maps are useful for identifying where additional source auditing and locally grounded research are needed, but not for ranking district safety or allocating services.

\newcommand{\districtmapfigure}{%
\begin{figure}[tbp]
\centering
\includegraphics[width=\linewidth]{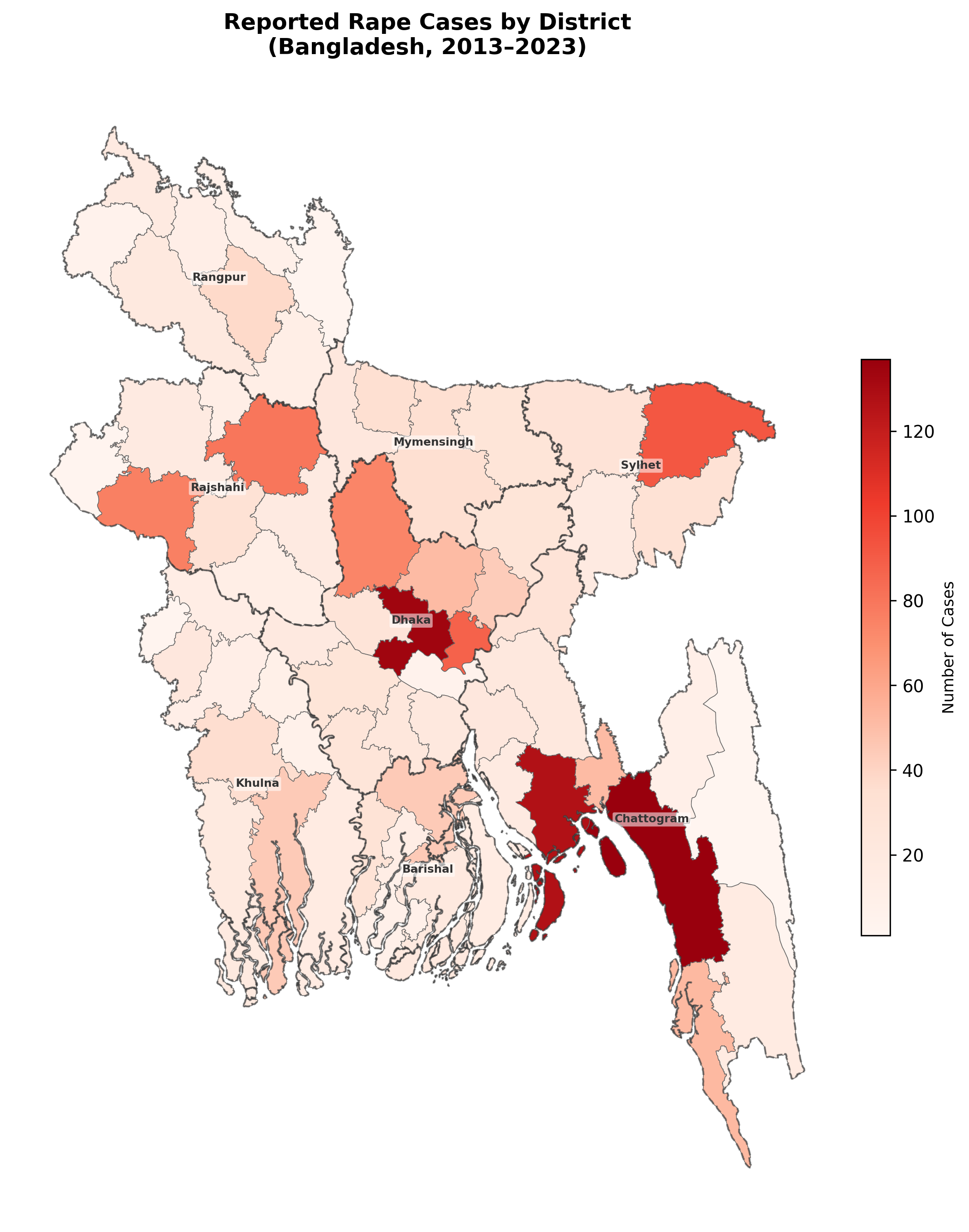}
\caption{Distribution of district-tagged article records. The 751 parsed records without a usable district tag are not represented. Raw counts indicate documentation density and are neither population-adjusted incidence nor risk.}
\Description{A choropleth map of Bangladesh at district level. Darker shading denotes more articles with a matching district tag.}
\label{fig:district-appendix}
\end{figure}
}

\subsection{Perpetrator Profiles}
One alleged perpetrator is recorded in 1,216 articles and more than one in 1,239. The metadata also contain 5,199 attacker-detail entries. Among specified relationships, acquaintance (1,410), stranger (1,315), and neighbor (854) dominate (Fig.~\ref{fig:case-patterns}c). Because these are extracted entries attached to articles, they are neither verified identities nor counts of unique people. Still, the mixture of stranger and socially proximate categories cautions against designing prevention only around stranger-danger scenarios.

Alleged-perpetrator age is available for 2,776 detail entries, with a median of 27 and mean of 29.8 years; the largest recorded age ranges are 25--34 (1,055 entries) and 18--24 (878). Gender is recorded as male in 5,004 of 5,199 entries (96.2\%). Student is the largest specified occupation label (151), followed by a heterogeneous set of political, transport, education, and law-enforcement labels. These occupation values are too sparse and sensitive to support group-level risk claims, but they illustrate the variety of institutional positions represented in reporting. Table~\ref{tab:attacker-summary} presents the full alleged-perpetrator profile directly in the findings, including completeness and privacy-sensitive fields.

\newcommand{\attackersummarytable}{%
\begin{table}[tbp]
\centering
\caption{Alleged-perpetrator characteristics in the parsed case metadata.}
\label{tab:attacker-summary}
\footnotesize
\renewcommand{\arraystretch}{1.18}
\begin{tabular}{@{}>{\centering\arraybackslash}m{0.34\linewidth}|>{\raggedright\arraybackslash}m{0.58\linewidth}@{}}
\toprule
\textbf{Field} & \textbf{Result} \\
\midrule
\textbf{Case-level attacker count} &
One attacker: 1,216 cases.\newline
Two attackers: 312 cases.\newline
Three attackers: 239 cases.\newline
Four attackers: 212 cases.\newline
Five or more: 476 cases.\newline
Unknown: 339 cases. \\
\midrule
\textbf{Attacker detail records} &
5,199 attacker-detail entries were extracted from 2,794 parsed case records. \\
\midrule
\textbf{Attacker age} &
Available: 2,776 attacker entries.\newline
Median: 27 years.\newline
Mean: 29.8 years.\newline
Age 18--24: 878 entries.\newline
Age 25--34: 1,055 entries. \\
\midrule
\textbf{Attacker gender} &
Male: 5,004 (96.2\%).\newline
Female: 115 (2.2\%).\newline
Unknown: 80 (1.5\%). \\
\midrule
\textbf{Relationship to victim} &
Acquaintance: 1,410.\newline
Stranger: 1,315.\newline
Neighbor: 854.\newline
Family member: 158.\newline
Unknown: 1,462. \\
\midrule
\textbf{Occupation} &
Known occupation: 1,664 entries.\newline
Unknown: 3,535 entries.\newline
Student: 151.\newline
Chhatra League worker: 95.\newline
Police officer: 48.\newline
Teacher: 42.\newline
Bus/vehicle driver: 82. \\
\midrule
\textbf{Identifying fields} &
Name recorded: 4,471 entries.\newline
Alias recorded: 435 entries.\newline
Residence recorded: 2,218 entries.\newline
Values are excluded from aggregate reporting for privacy. \\
\bottomrule
\end{tabular}
\end{table}
}

\attackersummarytable

\subsection{Incident Context}
An incident date is recorded in 2,055 articles (73.6\%), while a time is recorded in 631 (22.6\%) and a broader time/hour field in 1,401 (50.1\%). Incident location and descriptive detail are recorded in 2,695 (96.5\%) and 2,784 (99.6\%) articles, respectively, but exact values are excluded from aggregate presentation for privacy. Contextual indicators such as alcohol/drug involvement and fatal-weapon use are less consistently specified; their full availability is documented in the appendix, and an omitted mention is not treated as evidence of absence.

\newcommand{\incidentsummarytable}{%
\begin{table}[tbp]
\centering
\caption{Incident characteristics in the parsed case metadata.}
\label{tab:incident-summary}
\footnotesize
\renewcommand{\arraystretch}{1.18}
\begin{tabular}{@{}>{\centering\arraybackslash}m{0.34\linewidth}|>{\raggedright\arraybackslash}m{0.58\linewidth}@{}}
\toprule
\textbf{Field} & \textbf{Result} \\
\midrule
\textbf{Incident records} &
2,794 parsed incident records. \\
\midrule
\textbf{Temporal fields} &
Date recorded: 2,055 (73.6\%).\newline
Time recorded: 631 (22.6\%).\newline
Time/hour recorded: 1,401 (50.1\%). \\
\midrule
\textbf{Location and details} &
Location recorded: 2,695 (96.5\%).\newline
Incident details recorded: 2,784 (99.6\%).\newline
Exact values are excluded from aggregate reporting for privacy. \\
\midrule
\textbf{Husband involvement} &
No: 2,431 (87.0\%).\newline
Yes: 52 (1.9\%).\newline
Unknown: 311 (11.1\%). \\
\midrule
\textbf{Alcohol or drugs involved} &
No: 721 (25.8\%).\newline
Yes: 109 (3.9\%).\newline
Unknown: 1,964 (70.3\%). \\
\midrule
\textbf{Fatal weapon used} &
No: 1,238 (44.3\%).\newline
Yes: 221 (7.9\%).\newline
Unknown: 1,335 (47.8\%). \\
\midrule
\textbf{Romantic relationship} &
No: 1,970 (70.5\%).\newline
Yes: 295 (10.6\%).\newline
Unknown: 529 (18.9\%). \\
\midrule
\textbf{Victim killed after rape} &
No: 2,233 (79.9\%).\newline
Yes: 309 (11.1\%).\newline
Unknown: 252 (9.0\%). \\
\bottomrule
\end{tabular}
\end{table}
}

\subsection{Legal Response}
\begin{figure*}[t]
\centering
\includegraphics[width=\textwidth]{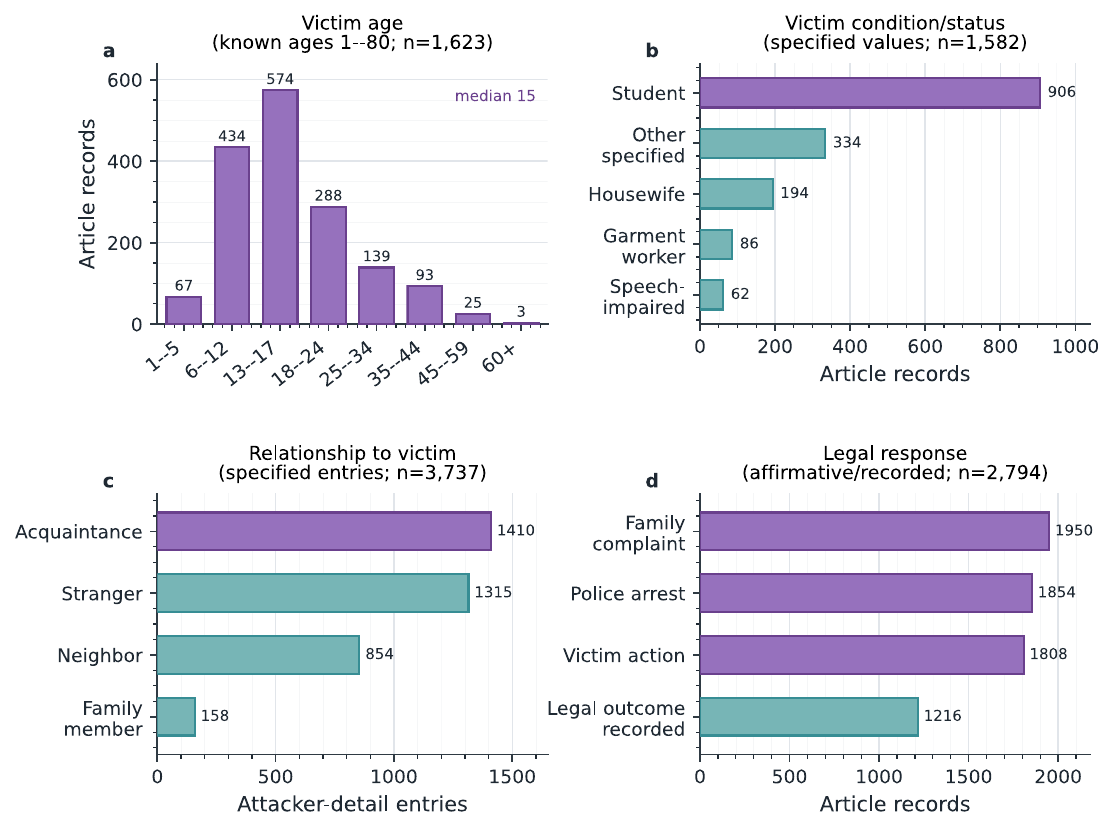}
\caption{Selected case-documentation patterns. (a) Victim age among 1,623 values within 1--80; three extracted values outside this range are omitted. (b) Specified victim condition/status labels ($n=1,582$); 1,212 omitted or unknown values are not plotted. (c) Specified relationship entries ($n=3,737$), which are neither unique people nor verified identities; 1,462 omitted or unknown values are not plotted. (d) Legal-response fields ordered by their plotted counts. Purple denotes affirmative immediate actions; teal distinguishes the availability of a recorded legal outcome, which is not a Yes/No action measure. Full denominators and completeness counts appear in Appendix~\ref{app:detailed-case-tables}. Counts characterize article metadata, not prevalence or verified case outcomes.}
\Description{Four panels. Panel a shows victim age groups, with ages 13 to 17 and 6 to 12 largest. Panel b shows specified victim condition or status labels, led by student. Panel c shows specified relationships to the victim, led by acquaintance and stranger. Panel d orders family complaint, police arrest, victim action, and recorded legal outcome by count. Immediate actions are purple, while legal-outcome availability is teal. Omitted or unknown categories are documented in the caption and appendix rather than plotted.}
\label{fig:case-patterns}
\end{figure*}

Legal-response fields contain more information about immediate actions than eventual outcomes. Police arrest is marked yes in 1,854 articles (66.4\%), victim action in 1,808 (64.7\%), and a family complaint in 1,950 (69.8\%). A legal outcome is recorded in 1,216 articles (43.5\%; Fig.~\ref{fig:case-patterns}d). These values describe what was available in an article at collection time, not final case disposition; the appendix reports the corresponding completeness audit.

The same asymmetry appears in the free-text fields. A police response is recorded in 2,217 articles and a police statement in 2,018, but a broader legal perspective is available in only 411. A reason for not filing a complaint is recorded in just 149 articles. Sparse mentions include fear, stigma, pressure, and financial constraints, but their frequencies cannot estimate the prevalence of reporting barriers. Overall, the metadata are better at documenting immediate police and complaint activity than tracing a case through prosecution and final disposition.

\newcommand{\legalsummarytable}{%
\begin{table}[tbp]
\centering
\caption{Legal-response patterns in the parsed case metadata.}
\label{tab:legal-summary}
\footnotesize
\renewcommand{\arraystretch}{1.18}
\begin{tabular}{@{}>{\centering\arraybackslash}m{0.34\linewidth}|>{\raggedright\arraybackslash}m{0.58\linewidth}@{}}
\toprule
\textbf{Field} & \textbf{Result} \\
\midrule
\textbf{Legal records} &
2,794 parsed legal records. \\
\midrule
\textbf{Police response text} &
Recorded: 2,217 (79.3\%).\newline
Unknown: 577 (20.7\%). \\
\midrule
\textbf{Police statement} &
Recorded: 2,018 (72.2\%).\newline
Unknown: 776 (27.8\%). \\
\midrule
\textbf{Police arrested criminal} &
Yes: 1,854 (66.4\%).\newline
No: 453 (16.2\%).\newline
Unknown: 487 (17.4\%). \\
\midrule
\textbf{Victim took action} &
Yes: 1,808 (64.7\%).\newline
No: 453 (16.2\%).\newline
Unknown: 533 (19.1\%). \\
\midrule
\textbf{Family filed complaint} &
Yes: 1,950 (69.8\%).\newline
No: 143 (5.1\%).\newline
Unknown: 701 (25.1\%). \\
\midrule
\textbf{Legal outcome} &
Recorded: 1,216 (43.5\%).\newline
Unknown: 1,578 (56.5\%).\newline
Repeated labels include ongoing investigation, pending, sent to jail, arrested, and case filed. \\
\midrule
\textbf{Officer in charge} &
Recorded: 2,031 (72.7\%).\newline
Unknown: 763 (27.3\%).\newline
Names are excluded from aggregate reporting. \\
\midrule
\textbf{Legal perspective} &
Recorded: 411 (14.7\%).\newline
Unknown: 2,383 (85.3\%). \\
\midrule
\textbf{Reason for not filing complaint} &
Recorded: 149 (5.3\%).\newline
Unknown: 2,645 (94.7\%).\newline
Sparse responses include fear, stigma, pressure, and financial constraints. \\
\bottomrule
\end{tabular}
\end{table}
}

The case-level community-response fields are much sparser than the incident and legal fields, so their detailed completeness summary is reported in Appendix~\ref{app:case-community-metadata}.

\subsection{Survey Perspectives}
The survey contains 115 convenience-sample responses collected from November 22 to December 9, 2020. The full descriptive summary, including response timing and closed-item distributions, is in Appendix~\ref{app:survey-detail}.

\newcommand{\surveyfulltable}{%
\begin{table}[tbp]
\centering
\caption{Detailed summary of survey responses.}
\label{tab:survey-summary}
\footnotesize
\renewcommand{\arraystretch}{1.18}
\begin{tabular}{@{}>{\centering\arraybackslash}m{0.34\linewidth}|>{\raggedright\arraybackslash}m{0.58\linewidth}@{}}
\toprule
\textbf{Field} & \textbf{Result} \\
\midrule
\textbf{Total responses} &
115 submitted responses. \\
\midrule
\textbf{Response timeline} &
November 22--December 9, 2020.\newline
November 26: 53 responses.\newline
November 23: 47 responses. \\
\midrule
\textbf{Respondent gender} &
Male: 61 (53.0\%).\newline
Female: 51 (44.3\%).\newline
Unspecified: 3 (2.6\%). \\
\midrule
\textbf{Rape-free society possible} &
Yes: 94 (81.7\%).\newline
No: 17 (14.8\%).\newline
Unclear or missing: 4 (3.5\%). \\
\midrule
\textbf{Abuse or unpleasant situation nearby} &
Yes: 65 (56.5\%).\newline
No: 47 (40.9\%).\newline
Missing: 3 (2.6\%). \\
\midrule
\textbf{Main perceived causes} &
Social degradation: 87.\newline
Unemployment: 3.\newline
Sadness: 2. \\
\midrule
\textbf{Why girls become victims} &
Lack of social security: 41.\newline
Social learning: 36.\newline
Patriarchal society: 14. \\
\midrule
\textbf{Female family members} &
2 members: 36 responses.\newline
3 members: 24 responses.\newline
4 members: 19 responses.\newline
1 member: 16 responses.\newline
Median: 3; range: 1--8. \\
\bottomrule
\end{tabular}
\end{table}
}

The respondent group was close to gender-balanced: 61 identified as male, 51 as female, and three were unspecified. Responses were collected over 18 days, but 100 of 115 submissions arrived on November 23 or 26. This concentration suggests clustered recruitment or circulation and further limits any claim that the survey represents public opinion over time or across Bangladesh.

Most respondents believed a rape-free society was possible (94/115), while 65/115 reported abuse or an unpleasant situation nearby. In the questionnaire's supplied categories, social degradation was selected 87 times as a perceived cause. Lack of social security (41), social learning (36), and patriarchal society (14) were the most frequent answers to the victimization question. These values show which explanations were salient within the instrument and respondent group; they are not tests of causal mechanisms.

\begin{figure}[htbp]
\centerline{\includegraphics[width=\linewidth]{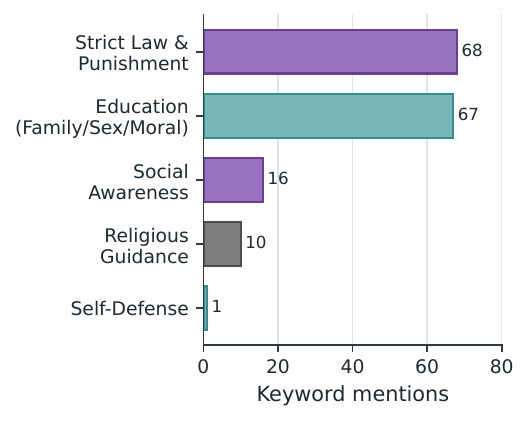}}
\caption{Provisional keyword mentions in open-ended prevention responses. Categories can overlap within a response and should not be interpreted as population prevalence or as a validated thematic model.}
\Description{A horizontal bar chart showing that law and punishment and education-related terms appear more often than social awareness, religious guidance, and self-defense terms in the open-ended responses.}
\label{fig:prevention}
\end{figure}

The open-ended prevention responses repeatedly connect legal enforcement with education, gender socialization in families, women's safety, and cultural change. The provisional keyword view in Fig.~\ref{fig:prevention} shows near-equal mentions of law/punishment and education-related terms, followed by social awareness and religious guidance. Counts are keyword mentions rather than mutually exclusive respondents, so one answer may contribute to multiple categories. The action-oriented question adds bystander resistance, survivor support, judicial accountability, and family responsibility. Section~\ref{sec:community-response} develops these patterns through the two full qualitative subsections and selected respondent excerpts.

Together, the streams show an interpretive tension rather than causal triangulation. The articles make adolescent and student victims, socially proximate alleged perpetrators, and incomplete legal outcomes visible; respondents emphasize enforcement, security, gender socialization, education, and community action. Their juxtaposition motivates design questions, but it cannot verify respondents' explanations or justify geographically targeted intervention from raw article counts. The next dedicated community-response section preserves the fuller interpretive account rather than reducing these responses to the keyword chart alone.

\newcommand{\communityresponsemain}{%
\section{Community Response}
\label{sec:community-response}

To complement the reported case metadata and spatial analysis, we analysed two open-ended survey questions from the Bangladesh Rape Case Survey. The preceding survey profile shows that most respondents believe prevention is possible, more than half report knowing someone who has faced abuse or an unpleasant situation, and the dominant closed-ended explanations emphasize social degradation, lack of social security, social learning, and patriarchal structure. The community-response analysis below interprets the open-ended answers behind those summary patterns; Appendix~\ref{app:survey-detail} retains their full descriptive context.

The first open-ended question asked respondents what should be done about rape, while the second asked how rape could be prevented. The two questions capture different but related dimensions of public perception. The first question produced more action-oriented responses, often framed around individual, family, and community responsibility. The second question produced broader prevention-oriented responses, focused on law, education, family socialisation, and cultural change. Together, these responses help explain how community members interpret both the causes of rape and the kinds of interventions they consider necessary.

\subsection{Community Action}

The question ``From your perspective, what do you think should be done about this?'' generated 82 responses, of which 80 were usable for thematic analysis. Compared with the prevention question, these responses were more personal and action-oriented. Many respondents used first-person language and described what they, their families, or their communities should do in response to rape and sexual harassment. Three major response groups were most important: community resistance, judicial accountability, and family-level moral correction.

\textbf{Community Mobilization.}
The strongest response group emphasized protest, awareness, and direct intervention. Respondents argued that ordinary citizens should not remain silent when they witness harassment, abuse, or misogyny. One male respondent described a sequence of action:

\begin{quote}
``The first thing to do is to be aware. Keep a watchful eye on whether anyone around me is being raped or sexually harassed. If such an incident has happened, then first of all, protest it and stand by the victim, demanding fair trial for the perpetrator. And if such a thing has never happened near me, then make others aware of it.''
\end{quote}

This response is important because it combines vigilance, protest, survivor support, and awareness-building. A female respondent more directly identified silence as part of the problem:

\begin{quote}
``Protest if you see any form of abuse/\allowbreak violence/\allowbreak misogyny against women. Because it is through silent tolerance that such moral degradation/\allowbreak mental distortion has taken such a large scale.''
\end{quote}

Together, these responses frame rape prevention as a community responsibility. Silence is not treated as neutral; rather, it is understood as a condition that allows violence and harassment to continue.

\textbf{Institutional Accountability.}
A second important response group focused on legal and institutional reform. Respondents called for exemplary trials, proper enforcement, speedy justice, and a justice system free from political or financial influence. One female respondent wrote:

\begin{quote}
``To conduct exemplary trials. To keep the judicial system free from party influence and financial influence. To keep it free from religious dogma. To fully ratify and implement the CEDO Charter. To reform the education system. To formulate a science-based education system.''
\end{quote}

This response is significant because it does not reduce justice to punishment alone. It identifies party influence, financial influence, religious pressure, education reform, and international rights frameworks as connected parts of the same structural problem. This aligns with the legal-response limitations observed in the case metadata, where many outcomes are unknown, pending, or incomplete.

\textbf{Family Education.}
A third group of responses located action inside the family. Respondents described teaching brothers, children, parents, and neighbours to respect women and challenge harmful attitudes. One female respondent wrote:

\begin{quote}
``We all have to be aware of this from our own families. We have to make changes ourselves first. If I see that the elders are presenting something wrongly to the children or are themselves having misconceptions about something, I immediately try to correct them.''
\end{quote}

Another respondent connected family education with survivor support:

\begin{quote}
``Teaching my children morality and accountability. Trying to make them better human rather than materialistic money machines. Also providing support to a victim and stand by her as per my abilities.''
\end{quote}

These responses suggest that participants see prevention as beginning in everyday social life. Families are not treated as private spaces outside the problem; instead, they are presented as key sites where harmful gender norms can either be reproduced or challenged.

\textbf{Women’s Empowerment.}
Several female respondents also emphasized self-defence, mental strength, and the ability to seek help without shame. One respondent wrote:

\begin{quote}
``Every woman needs to become mentally strong, take self-defense training, and seek legal help without shame/fear.''
\end{quote}

This response should not be read as shifting responsibility onto women. Rather, it reflects a context where many women feel that legal and social systems may not protect them quickly enough. Women’s self-protection is therefore imagined as one part of a broader strategy that also requires legal reform and community support.

\subsection{Prevention Pathways}

The question ``How do you think it is possible to prevent rape?'' generated 98 responses, of which 93 English-language responses were usable for thematic analysis. These responses were broader and more policy-oriented than the previous question. Respondents identified rape prevention as a layered problem involving law, education, family socialisation, women’s safety, and cultural transformation. The most important response groups were legal reform, education reform, family-based gender socialisation, and social awareness.

\textbf{Legal Reform.}
The most frequent prevention group focused on law and justice. Respondents repeatedly mentioned strict law enforcement, exemplary punishment, removal of loopholes, and ending impunity. One female respondent gave one of the most structurally developed answers:

\begin{quote}
``Full ratification and implementation of the CEDAW Charter, introduction of a non-discriminatory, science-based education system. Exemplary punishment of rapists. Elimination of impunity for crimes. Ending party patronage of criminals.''
\end{quote}

This response connects rape prevention with international legal commitments, science-based education, impunity, and political patronage. Another female respondent similarly emphasized:

\begin{quote}
``Through strict enforcement of the justice system, without leaving loopholes in the law, and by changing the perspective of all people in society when women protest.''
\end{quote}

These responses show that participants understand rape as more than an individual crime. They see it as a problem sustained by weak legal accountability, social backlash against women, and protection of powerful perpetrators.

\textbf{Preventive Education.}
A second major group emphasized education. Respondents called for moral education, social education, sex education, and reform of the national education system. One female respondent summarized this link clearly:

\begin{quote}
``Sexual education, change in attitudes towards women, improvement of women's social status, proper implementation of laws.''
\end{quote}

This response is concise but important because it places sex education, women’s social status, and law enforcement inside the same prevention framework. Another respondent argued that prevention requires structural educational change:

\begin{quote}
``It is possible to prevent rape by developing the practice of morality and human values in our families, society, and state education systems. Structural changes in the education system are needed.''
\end{quote}

These responses suggest that respondents do not see education only as schooling. They understand it as a wider process through which society teaches respect, sexuality, responsibility, and gender relations.

\textbf{Gender Socialisation.}
A third important group focused on the family as the first place where unequal gender norms are produced. Several respondents criticized the tendency to restrict girls while failing to correct boys. One female respondent wrote:

\begin{quote}
``Keeping in mind the safety of a girl by imposing restrictions on her movement, have we been able to correct those for whom this is happening? In the family, boys are given freedom at a very young age, it is important to pay attention to whether they are abusing it or not.''
\end{quote}

This response shifts the burden of prevention away from girls and toward boys, families, and early socialisation. Another respondent connected prevention to childhood gender roles:

\begin{quote}
``We should not give only dolls and pots and pans to girls and only footballs and bullets to boys... From the beginning, we should give the same discipline to the children. Both children should be taught to respect each other.''
\end{quote}

These responses show a strong awareness that gender inequality is learned early. Prevention, therefore, must begin before violence occurs: in the way children are raised, disciplined, and taught to understand each other.

\textbf{Cultural Change.}
A final group emphasized social awareness, cultural development, and sustained monitoring of the problem. One female respondent described prevention as a long-term process of diagnosis, action, and evaluation:

\begin{quote}
``Finding out the problem, why rape is happening, who is doing it, what is their problem, why we are not able to save ourselves from rape. Then finding a way to solve the problem, working on it for a specific time. And constantly monitoring the pace of work and results.''
\end{quote}

This response closely reflects the logic of the present study. It calls for identifying causes, understanding perpetrators, examining why prevention fails, designing solutions, and monitoring progress. This suggests that some respondents already view rape prevention as an evidence-based social process rather than a one-time punitive response.

Overall, the two survey questions reveal a layered public understanding of rape and its prevention. The action-oriented responses emphasize protest, survivor support, family correction, and personal responsibility. The prevention-oriented responses emphasize law, education, gender socialisation, and cultural transformation. Taken together, they suggest that rape prevention in Bangladesh requires more than legal punishment alone. It requires coordinated action across institutions, families, schools, communities, and everyday social norms.
}

\communityresponsemain

\section{Discussion}
The evaluation suggests that sexual violence in Bangladesh should be analysed as a layered social problem rather than a single-factor crime pattern. The articles make female, young, and student victims especially visible, while socially proximate relationship categories, taken together, outnumber stranger entries. This documentation pattern complicates prevention narratives that focus only on public-space danger and motivates attention to household, neighborhood, educational, and community contexts as well.

Spatially, the district and subdistrict maps show that documentation concentration is uneven. This may reflect local variation, media coverage, reporting behavior, population density, internet visibility, acquisition choices, or tagging error. High-count areas can motivate reporting audits and locally grounded follow-up, but the maps cannot rank safety or determine resource priority. Conversely, low counts may reflect missing or suppressed reporting rather than safety.

The community responses clarify how participants interpret these documentation patterns. Respondents do not describe rape prevention as only a matter of punishment, even though strict legal enforcement is a dominant theme. They repeatedly connect legal accountability with education reform, family-based socialization, social awareness, women's safety, survivor support, and public resistance to harassment. Placing this discussion after the full community-response analysis allows those qualitative findings to inform the interpretation rather than appearing as a conclusion before the evidence is presented.

\subsection{Design Implications}
These interpretations translate into concrete HCI requirements and research priorities rather than direct resource-allocation recommendations. First, public-facing legal information systems could make arrest, filing, prosecution status, and survivor-support pathways easier to follow, while protecting identities. Second, adolescent- and student-centered prevention deserves evaluation because these groups are prominent among records with known fields. Third, tools should not encode only stranger-danger scenarios; the corpus also documents acquaintance and neighborhood contexts. Fourth, any operational geographic system must combine population-adjusted official data, reporting audits, source provenance, and uncertainty displays before it is used to allocate resources.

The survey responses also point toward practical community-level programming. Respondents emphasize legal reform, education, social awareness, and bystander action, suggesting that public campaigns should move beyond general slogans and provide concrete guidance: how to report, how to support survivors, how to intervene safely, and how families and schools can challenge rape-supportive attitudes.

\section{Limitations}
Several limitations affect interpretation. The case dataset is drawn from one publishing ecosystem and is subject to selection bias: incidents that receive Prothom Alo coverage are more likely to enter the corpus, while unreported, locally suppressed, or differently framed incidents may be absent. The acquisition query, inclusion/exclusion process, acquisition date, and deduplication procedure are not preserved in the repository. The early years contain almost no records, 2023 may be incomplete, and location tags are frequently missing. The article is the current unit of analysis, so follow-up stories may duplicate incidents. Raw geographic counts lack population denominators and a model of news/reporting access.

The structured metadata are an additional measurement layer. The extraction system, prompt/schema provenance, translation procedure, and validation annotations are not available, so field-level accuracy and bias are unknown. Categories can reflect newsroom wording and extraction error as well as events. In particular, missing article detail cannot be interpreted as ``No,'' and sensitive labels such as relationship, occupation, political affiliation, alcohol/drug involvement, and legal outcome require manual validation.

The survey dataset is also limited. It contains 115 convenience-sample responses and should not be treated as nationally representative. Recruitment, location, eligibility, consent, and translation validation are undocumented. The question asking for ``the reason for the increase'' presupposes that an increase occurred, and the question ``Why are girls victims?'' can invite victim-focused framing; these are instrument-design threats. The current grouping of open responses lacks a documented codebook, multiple coders, agreement process, and reflexivity statement. Keyword frequencies are descriptive signals, not statistically generalizable themes.

\section{Future Work}
Future work should extend the dataset with additional years, sources, and validation procedures. Linking reported cases to population, poverty, education, transport, urbanization, and law-enforcement indicators would allow more robust spatial modeling. District-level and subdistrict-level rates should be normalized by population and reporting access. A larger representative survey could test whether the perception patterns observed here hold across gender, age, class, education, region, and rural--urban location. Finally, future research should evaluate prevention interventions directly, especially school-based consent education, survivor reporting support, bystander training, and low-connectivity digital emergency tools.

\section{Conclusion}
This paper combines structured metadata from a Prothom Alo--ecosystem article corpus with a small community survey. The corpus documents many young, female, and student victims, socially proximate alleged perpetrators, uneven geographic documentation, and incomplete legal-response information. Survey respondents describe social degradation, weak enforcement, insecurity, patriarchal norms, education gaps, and community inaction as perceived enabling conditions.

The paper's central contribution is an uncertainty-aware integration of case-documentation patterns with perception data. The analysis demonstrates both the value and danger of making fragmented reports computationally legible: structured views can reveal recurring documentation patterns, but they can also turn source bias and missingness into false claims of prevalence, risk, or cause. For HCI, responsible systems in this domain must make provenance, uncertainty, privacy, and validation visible while supporting legal transparency, education, survivor support, and careful local inquiry.

\appendix
\section{Metadata Tables}
\label{app:detailed-case-tables}
The main text now contains the complete victim and alleged-perpetrator profile tables because those distributions are central to the findings. This appendix retains the supporting coverage, incident, legal, survey, community, and geographic audits. Table~\ref{tab:field-coverage} first records the availability of fields that materially shape interpretation; Tables~\ref{tab:incident-summary} and~\ref{tab:legal-summary} then preserve supporting contextual and legal-response detail.

The default unit is one parsed article record. Rows explicitly labeled as attacker-detail entries instead describe nested people mentioned within articles; these entries are not verified identities and may not represent unique individuals. Likewise, an article is not necessarily a unique incident because follow-up or duplicate reporting has not yet been fully linked. Percentages therefore describe the acquired and parsed corpus, not national prevalence, incidence, or population risk.

``Unknown'' combines fields absent from the source representation with values explicitly encoded as unknown. It is reported here to make denominator changes visible, but it is not treated as a negative response. Names, exact residences, detailed incident locations, officer names, and case narratives are excluded even when present because aggregate analysis does not require publishing potentially identifying information. All tables inherit uncertainty from source selection, article acquisition, and automated metadata extraction; a stratified manual-validation study remains necessary before inferential analysis.

Table~\ref{tab:field-coverage} is a coverage audit rather than a result ranking. It shows why the age, condition/status, geography, and legal analyses use different denominators. In particular, district coverage is substantially stronger than subdistrict coverage, and eventual legal outcomes are less frequently documented than immediate legal actions.

\fieldcoveragetable

Table~\ref{tab:incident-summary} distinguishes frequently recorded narrative fields from less consistently specified contextual indicators. The table also documents why exact locations and incident narratives are excluded: their analytical value here does not outweigh their re-identification risk.

\incidentsummarytable

Table~\ref{tab:legal-summary} shows the difference between immediate procedural reporting and longer-term case tracking. Arrests, complaints, and police statements are commonly recorded, whereas legal perspective, reasons for not filing, and final outcomes are much less available. This gap motivates longitudinal linkage rather than treating an unreported outcome as no legal action.

\legalsummarytable
\FloatBarrier

\section{District Map}
\label{app:district-map}
The main text presents the finer subdistrict view to show within-district documentation patterns. Figure~\ref{fig:district-appendix} provides the complementary district-level audit, using the 2,043 parsed articles with a usable district tag; 751 parsed records do not appear. Compared with the subdistrict view, this map includes a larger share of the corpus and is therefore useful for checking whether fine-grained concentrations persist after aggregation.

The map uses raw documentation counts. Darker areas may reflect population size, newsroom reach, reporting practices, source visibility, place-name normalization, or acquisition decisions in addition to any underlying variation in violence. The figure should therefore guide source auditing and local follow-up rather than district ranking, risk prediction, or resource allocation. A future spatial analysis would require deduplicated incidents, population denominators, validated geographic joins, and uncertainty estimates.

\districtmapfigure
\FloatBarrier

\section{Survey Details}
\label{app:survey-detail}
Table~\ref{tab:survey-summary} retains the response timeline, full closed-item summary, and household descriptor omitted from the main text. The 115 responses form a convenience sample collected over 18 days in late 2020, and 100 submissions arrived on only two dates. This clustering is consistent with concentrated recruitment or circulation and prevents the sample from representing Bangladeshi public opinion.

Closed-item counts describe only the supplied questionnaire categories. Open-ended answers were provisionally grouped through repeated concepts and keyword mentions; categories can overlap, and a single answer may contribute to several counts. The present repository does not include a finalized qualitative codebook, independent double-coding, disagreement resolution, or complete recruitment and consent documentation. Consequently, the survey supports an exploratory account of respondent perspectives, not population inference or causal explanation.

\surveyfulltable
\FloatBarrier

Table~\ref{tab:survey-summary} shows that belief in the possibility of prevention coexists with substantial reported proximity to abuse or unpleasant situations. Because recruitment was concentrated on two dates and the questionnaire supplied several answer categories, the table contextualizes the qualitative responses but cannot establish the distribution of beliefs in Bangladesh.

\section{Community Metadata}
\label{app:case-community-metadata}
Table~\ref{tab:case-community-summary} summarizes the case-level community fields extracted from the media metadata. These fields are used mainly to identify whether public response, pressure, or local arbitration was documented in the source article. They are distinct from the 2020 respondent survey: this table describes what journalists recorded about a reported case, whereas Section~\ref{sec:community-response} examines what survey participants proposed.

Because these fields are sparsely populated, the counts should not be interpreted as the frequency of community protest, family pressure, or local arbitration in Bangladesh. An absent value can mean that an event did not occur, was not mentioned, was not available when the article was written, or was not recovered by extraction. The detailed table is retained because these gaps identify priorities for future annotation and longitudinal follow-up. Names of local leaders and case-specific accounts remain excluded to reduce re-identification risk.

\begin{table}[tbp]
\centering
\caption{Community-response fields in the parsed case metadata.}
\label{tab:case-community-summary}
\footnotesize
\renewcommand{\arraystretch}{1.18}
\begin{tabular}{@{}>{\centering\arraybackslash}m{0.34\linewidth}|>{\raggedright\arraybackslash}m{0.58\linewidth}@{}}
\toprule
\textbf{Field} & \textbf{Result} \\
\midrule
\textbf{Community response text} &
Recorded: 479 (17.1\%).\newline
Unknown: 2,315 (82.9\%). \\
\midrule
\textbf{Public response text} &
Recorded: 353 (12.6\%).\newline
Unknown: 2,441 (87.4\%). \\
\midrule
\textbf{Pressure from attacker's family} &
Text recorded: 295 (10.6\%).\newline
Unknown text: 2,499 (89.4\%). \\
\midrule
\textbf{Pressure indicator} &
Yes: 302 (10.8\%).\newline
No: 99 (3.5\%).\newline
Unknown: 2,393 (85.6\%). \\
\midrule
\textbf{Local arbitration leaders} &
Recorded: 188 (6.7\%).\newline
Unknown: 2,606 (93.3\%).\newline
Names are excluded from aggregate reporting. \\
\midrule
\textbf{Arbitration outcome} &
Recorded: 142 (5.1\%).\newline
Unknown: 2,652 (94.9\%).\newline
Sparse outcomes include settlement, punishment, forced marriage, police handover, or no action. \\
\bottomrule
\end{tabular}
\end{table}

Table~\ref{tab:case-community-summary} is best understood as a reporting-availability table. Community response text appears in 479 articles, while pressure and arbitration fields occur in a much smaller subset. These values identify what future case annotation should follow more systematically; they do not show that community pressure or local arbitration was absent from the remaining incidents.

\end{document}